\documentclass[journal]{IEEEtran}

\usepackage{amsmath,amssymb,amstext} 
\usepackage{bbm}  
\usepackage{textcomp}  
\usepackage{bbold}  
\usepackage[pdftex]{graphicx}
\usepackage{caption}
\usepackage{multirow}
\usepackage{array}
\usepackage{float}  
\usepackage{url}
\usepackage{xspace}  
\usepackage{enumitem}  
\usepackage{authblk}  
\usepackage{rotating}
\usepackage{pdflscape}
\usepackage{gensymb}  
\usepackage[disable]{todonotes}  
\usepackage{changes}  

\renewcommand{\vec}[1]{\mathbf{#1}}

\newcommand{\petco}{P$_{\text{ET}}$CO$_2$}

\title{Optical Hemodynamic Imaging of Jugular Venous Dynamics During Altered Central Venous Pressure}

\author{Robert Amelard,~\IEEEmembership{Member,~IEEE,}
        Andrew D Robertson, Courtney A Patterson, Hannah Heigold, Essi Saarikoski, Richard L Hughson
\thanks{This work was supported by the Natural Sciences and Engineering Research Council of Canada (PDF-503038-2017) and Canada Space Agency (18FAWATA13). (\textit{R.~Amelard and A.~D.~Robertson contributed equally to this work. Corresponding author: R.~Amelard}).}
\thanks{R. Amelard is with the Schlegel-UW Research Institute for Aging, Waterloo, Canada (e-mail: ramelard@uwaterloo.ca).}
\thanks{R.~L. Hughson is with the Schlegel-UW Research Institute for Aging, Waterloo, Canada.}
\thanks{A.~D.~Robertson, C.~Patterson, and H.~Heigold are with the Schlegel-UW Research Institute for Aging, Waterloo, Canada, and University of Waterloo, Waterloo, Canada.}
\thanks{E.~Saarikoski is with the Medical Research Center Oulu, Research Unit of Biomedicine, Oulu University Hospital and University of Oulu, Oulu, Finland, and Schlegel-UW Research Institute for Aging, Waterloo, Canada.}}

\date{}

\begin{document}

\maketitle

\bstctlcite{IEEEexample:BSTcontrol}

\begin{abstract}
\textit{Objective:} An optical imaging system is proposed for quantitatively assessing jugular venous response to altered central venous pressure.
\textit{Methods:} The proposed system assesses sub-surface optical absorption changes from jugular venous waveforms with a spatial calibration procedure to normalize incident tissue illumination. Widefield frames of the right lateral neck were captured and calibrated using a novel flexible surface calibration method. A hemodynamic optical model was derived to quantify jugular venous optical attenuation (JVA) signals, and generate a spatial jugular venous pulsatility map. JVA was assessed in three cardiovascular protocols that altered central venous pressure: acute central hypovolemia (lower body negative pressure), venous congestion (head-down tilt), and impaired cardiac filling (Valsalva maneuver).
\textit{Results:} JVA waveforms exhibited biphasic wave properties consistent with jugular venous pulse dynamics when time-aligned with an electrocardiogram. JVA correlated strongly (median, interquartile range) with invasive central venous pressure during graded central hypovolemia (r=0.85, [0.72, 0.95]), graded venous congestion (r=0.94, [0.84, 0.99]), and impaired cardiac filling (r=0.94, [0.85, 0.99]).
Reduced JVA during graded acute hypovolemia was strongly correlated with reductions in stroke volume (SV) (r=0.85, [0.76, 0.92]) from baseline (SV: 79$\pm$15~mL, JVA: 0.56$\pm$0.10~a.u.) to \textminus40~mmHg suction (SV: 59$\pm$18~mL, JVA: 0.47$\pm$0.05~a.u.; p$<$0.01).
\textit{Conclusion:} The proposed non-contact optical imaging system demonstrated jugular venous dynamics consistent with invasive central venous monitoring during three protocols that altered central venous pressure.
\textit{Significance:} This system provides non-invasive monitoring of pressure-induced jugular venous dynamics in clinically relevant conditions where catheterization is traditionally required, enabling monitoring in non-surgical environments.

\end{abstract}

\section{Introduction}

The cardiovascular system is a closed-loop system in which cardiac output and venous return remain in balance under state-steady conditions. Adequate venous return is essential to maintain cardiac output and regulate arterial blood pressure. Venous return is largely determined by the pressure gradient between the systemic veins and the right atrium~\cite{guyton1955}. Due to the high compliance of veins, venous blood volumes provide pressure-related fluid biomarkers of cardiac, systemic, and cerebrovascular regulation~\cite{dupont2011,holmlund2018}.

Central venous pressure (CVP) is a major determinant of cardiac filling pressures and preload. Elevated jugular venous pressure (JVP) is a primary indicator of congestion in heart failure~\cite{aha2013}. Heart failure and other diseases that restrict venous drainage, such as jugular vein thrombosis, are primary causes of intracranial hypertension and papilledema~\cite{thandra2015,zhou2018}. Head-down tilt (HDT), a model for spaceflight microgravity and venous stagnation~\cite{zhang2018}, has demonstrated chronically elevated intracranial pressure above seated levels on Earth over 24~h~\cite{lawley2017icp}, and recent investigations found venous thrombosis and stagnation/retrograde flow in astronauts during spaceflight~\cite{marshallgoebel2019}. In contrast, hypovolemic states, such as acute blood loss, induce arterial pressure changes secondary to reduced cardiac filling with lower CVP~\cite{alian2014}. Monitoring changes in CVP may provide important biomarkers relevant to cardiac and cerebral function; however, this measurement is not routinely performed due to invasive surgical right heart catheterization required to assess these pressure changes.

Due to its proximity to the heart, the jugular vein hemodynamics closely reflect those of CVP in many conditions. Elevated JVP is a biomarker of congestive heart failure, and provides insight into patient fluid status and response to diuresis~\cite{thibodeau2018,vinayak2006}. Clinically, jugular venous assessment is a visual task prone to inter-rater reliability errors~\cite{brennan2007jvpphysical}. Ultrasound has been used to quantify jugular distension for evaluating heart failure and right atrial pressure~\cite{pellicori2015valsalvahf,simon2018valsalvahf}, but is not suitable for longitudinal continuous monitoring. Mobile health devices capable of tracking JVP may enable breakthrough heart failure management beyond patient-reported functional and physiological data~\cite{devore2019}.

There has been recent interest in non-invasive and non-contact optical jugular venous monitoring. Skin surface displacement assessment from video~\cite{moco2018,lampotang2018,abnousi2019} can enhance jugular venous pulse identification, but does not contain information related to jugular distension. Other methods have been derived from non-contact photoplethysmographic imaging systems that traditionally monitor arterial hemodynamics~\cite{sun2015review}. Photon migration properties of near-infrared light enable deeper tissue penetration than superficial visible spectrum light~\cite{jacques2013}. Thus, optical systems have been investigated to assess the JVP waveform using variations in optical absorption resulting from hemodynamic fluctuations in blood volume~\cite{garcialopez2020,amelard2017}. Since pressure changes induce jugular distension and relaxation, optical approaches are promising for quantifying changes in JVP.

In this paper, we propose an optical imaging system to assess jugular venous dynamics associated with altered CVP in a point-of-care setting. The main contributions of this work are: (1) an optical model for quantifying optical changes related to hemodynamic processes (i.e., vessel pulsatility), (2) a novel surface calibration method that forms to curved surfaces of the neck, and (3) an image processing pipeline for denoising and identifying the jugular venous pulse spatial distribution. The calibration procedure enables monitoring dynamic venous changes within an individual by normalizing tissue reflectance with varying camera-tissue geometry. We designed a cardiovascular protocol that modified CVP in three deterministic ways by controlling pressure gradients along the body: (1) reduced venous return from acute central hypovolemia through graded lower body negative pressure (LBNP); (2) increased CVP and venous congestion through head-down tilt (HDT); and (3) impaired cardiac filling from increased intrathoracic pressure through a guided Valsalva maneuver. Taken together, this protocol evaluated system response across clinically and physiologically relevant perturbations by simulating effects of hemorrhagic trauma, venous congestion, and heart failure.

\section{Jugular Optical Imaging}
\begin{figure*}
    \centering
    \includegraphics[width=0.65\textwidth]{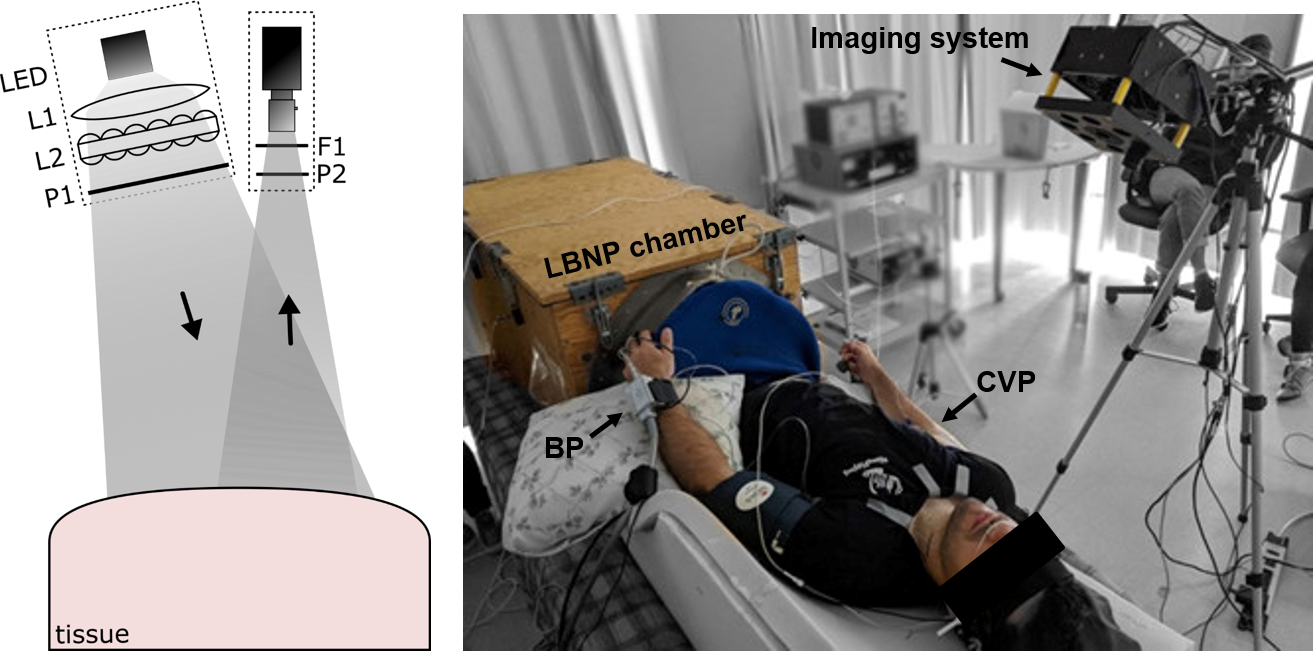}
    \caption{\textbf{Left:} diagram of the optical imaging instrumentation. LED illumination (940~nm) was passed through a condenser lens (L1), spatial homogenizer (L2), and linear polarizer (P1). Reflected light was cross polarized (P2) and near-infrared bandpass filtered (F1) prior to being imaged by the camera. \textbf{Right:} setup showing the lower-body negative pressure (LBNP) chamber, continuous blood pressure (BP), central venous pressure (CVP) line, and the imaging system overhead. Calibration was performed by placing a flexible reflectance target across the neck prior to image acquisition in the same position (see Fig.~\ref{fig:system_diagram}).}
    \label{fig:optical_setup}
\end{figure*}

\begin{figure*}
    \centering
    \includegraphics{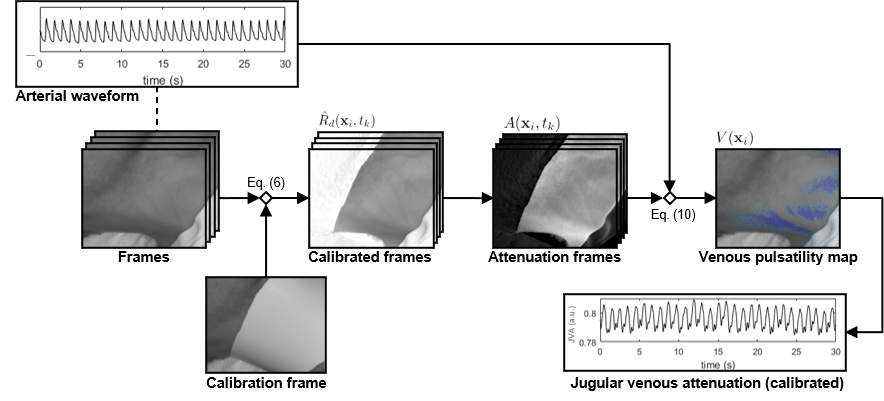}
    \caption{Overview of the jugular optical imaging system. Frames were captured of the right lateral area of the neck. Reflectance was calibrated using a custom flexible diffuse calibration target (see ``Calibration frames''). Attenuation frames were computed from motion corrected and denoised reflectance frames, and used with the arterial waveform to generate a venous pulsatility map for jugular waveform extraction.}
    \label{fig:system_diagram}
\end{figure*}

\subsection{Instrumentation}
Fig.~\ref{fig:optical_setup} depicts the optical imaging setup.
The tissue was illuminated with a 940~nm (FWHM 40~nm) LED (LZ1-10R702, LEDEngin) for increased tissue penetration due to low melanin absorption, epidermal and dermal scattering, and hemoglobin absorption~\cite{jacques2013}, thus increasing the probability of photons returning to the surface after penetrating the vessel. A uniform beam profile was attained through a mounted diffused spot beam lens optic and custom Fly's eye condenser, and polarized with a linear polarizer. These components were contained in a custom tube assembly and mounted in the same plane as the camera, beneath the camera lens. Tissue reflection was cross-polarized to reduce surface-level specular reflection, and near-infrared bandpass filtered prior to being imaged by the camera (GS3-U3-41C6NIR-C, FLIR).

\subsection{Optical Model}

For bulk homogeneous tissue, light attenuation (or optical density) can be calculated by the modified Beer-Lambert law~\cite{delpy1988}:
\begin{equation}
    A = -\log \left( \frac{I}{I_0} \right) = \sum_i \mu_{a,i} L + G
    \label{eq:mbll}
\end{equation}
where $A$ is attenuation, $I$ and $I_0$ are the detected and incident illumination, $\mu_a=\log(10)\varepsilon_i c_i$ is the absorption coefficient of chromophore $i$ based on its molar extinction coefficient ($\varepsilon_i$) and concentration ($c_i$), $L$ is the photon mean path length through the tissue, and $G$ is an intensity loss term due to scatter. With pulsatile flow, we are interested in modeling the temporal dynamics of light attenuation due to changes in blood volume. A standard approach in continuous wave near-infrared spectroscopy is to use the attenuation differential with multiple wavelengths to estimate changes in hemoglobin concentrations~\cite{scholkmann2014}. However, when considering pulsatile vessels in unchanging tissue, it is primarily the photon path length that changes with vessel expansion/contraction~\cite{shelley2007}. Thus, expressing Eq.~(\ref{eq:mbll}) in terms of time in this pulsatile model yields:
\begin{equation}
    A(t) = -\log \left( \frac{I(t)}{I_0} \right) = \mu_{a,t} L_t + \mu_{a,v} L_v(t) + G
    \label{eq:At}
\end{equation}
where $\mu_{a,t}$ and $\mu_{a,v}$ are the absorption coefficients for the surrounding tissue and vessel, respectively, similarly for $L_t$ and $L_v$. We can further expand the vessel absorption:
\begin{equation}
    \mu_{a,v} = \alpha \mu_{a,HbO_2} + (1-\alpha) \mu_{a,Hb}
\end{equation}
where $\alpha$ is the blood oxygen saturation in the vessel. Expressing attenuation in terms of temporal differences yields:
\begin{equation}
    \Delta A = -\log \left( \frac{I(t_2)}{I(t_1)} \right) = \mu_{a,v}(L_v(t_2) - L_v(t_1))
    \label{eq:deltaA}
\end{equation}
Thus, assuming constant oxygen saturation in the vessel (constant $\alpha$), and that changes in the bulk tissue properties are small compared to changes in vessel absorption, changes in measured attenuation are linearly proportional to changes in mean photon path length through the vessel resulting from changes in vessel blood volume.

\subsection{Imaging Calibration}
The formulation of Eq.~(\ref{eq:deltaA}) only holds if $I_0$ is constant across the differential measurements.
Changes in tissue-imaging orientation cause differences in illumination profiles, making cross-condition comparison unreliable, since $I_0$ becomes a function of camera-tissue orientation. To compensate for this effect, the imaging field was calibrated prior to each acquisition by deriving a widefield reflectance normalization term using planar light from the more general spatial frequency domain imaging~\cite{cuccia2009} by solving for when $f_x=0~\text{mm}^{-1}$:
\begin{align}
    R_d(\vec{x}_i) &= \frac{M_{tissue}(\vec{x}_i)}{M_{ref}(\vec{x}_i)} \hat{R}_{d,ref}(\vec{x}_i) \\
    M(\vec{x}_i) &= I_0(\vec{x}_i) \cdot \text{MTF}_{system}(\vec{x}_i) \cdot R_d(\vec{x}_i)
\end{align}
where $R_d(\vec{x}_i)$ is the diffuse reflectance at location $\vec{x}_i$, $\hat{R}_{d,ref}$ is the modeled reflectance profile of the calibration target, and $M$ is the reflected photon density wave for the tissue ($M_{tissue}$) and calibration reference target ($M_{ref}$), which is a function of the incident illumination ($I_0$), modulation transfer function of the optical system ($\text{MTF}_{system}$), and the true diffuse reflectance of the turbid medium ($R_d(\vec{x}_i)$). This formulation was used to compute calibrated diffuse reflectance signals $R_d(\vec{x}_i)$ by calibrating both the illumination source and optical system simultaneously with a single acquisition using a known calibration target. The calibration target must be thin and conform to the profile of the tissue for the incident illumination terms to cancel each other out (i.e., for $I_0$ in $M_{tissue}$ and $M_{ref}$ to be equivalent). Under these conditions, this calibration procedure corrects for surface profilometry and illumination inhomogeneities. 

Fig.~\ref{fig:system_diagram} depicts the processing pipeline for extracting jugular waveforms from a set of optical images. Reference images were acquired using a custom flexible calibration target that was wrapped around the participant's neck such that it followed the contour of the tissue.
The calibration target was constructed from a 1/16~in white nitrile rubber sheet. Its strong scattering properties resulted in minimal transmittance through the thin medium and maximal reflectance, ensuring consistent spatial and temporal properties that are unaffected by underlying tissue. The material's resilience at thin profiles allowed it to conform to the tissue profile across different body habitus. Two thousand grit sandpaper was used to produce a highly diffuse uniform reflectance profile and minimize surface specular reflectance, thus approximating a Lambertian surface providing uniform reflectance independent of the camera's angle of view.
The predicted calibration reflectance profile $\hat{R}_{d,ref}$ was therefore modeled as a uniform distribution at unity, describing uniform diffuse reflectance of the surface.
Calibration geometry was designed to conform to the participant's neck with no gap between the tissue and target, extending from the clavicle to the mandible with good lateral coverage. Three sizes of the calibration target were developed to account for anthropometric differences across participants. The calibration target was secured to the tissue using surgical tape, which was applied to the back of the calibration target outside the imaging area of interest.
A calibration image was obtained before every acquisition by averaging 120~frames of the calibration target wrapped around the participant's neck.

\subsection{Venous Pulsatility Map}
The diffuse reflectance map $R_d$ was down-sampled using a 4x4 mean filter to increase signal-to-noise ratio, and then each diffuse reflectance signal $R_d(\vec{x}_i)$ was denoised using a hemodynamic Kalman filter formulation~\cite{amelard2015}.
This temporal domain filter has been shown to preserve waveform shape and feature timings by incorporating smoothness priors and system-specific noise models into a linear state estimation system conditioned on previous states. Briefly, the model was defined as a linear system:
\begin{align}
    \vec{s}_k &= A\vec{s}_{k-1} + \vec{w}, \\
    \vec{z}_k &= H\vec{s}_k + \vec{v}
\end{align}
where $\vec{s}_k$ and $\vec{z}_k$ are the observed and estimated states at time $k$ based on the state transition matrix $A$ and observation matrix $H$, under the influence of process and measurement noise ($\vec{w}$ and $\vec{v}$ respectively). The smooth biophysical properties of pulsatile blood flow were incorporated into the model by defining the system state at each time point $k$ according to the signal measurement and rate of change, and the denoised signal was estimated using a smooth continuous vessel wall motion prior:
\begin{equation}
\vec{s}_k = \begin{bmatrix}
    s_k \\
    \dot{s}_k
  \end{bmatrix} = \begin{bmatrix}
    1 & \Delta t \\
    0 & 1
  \end{bmatrix} \vec{s}_{k-1}
\end{equation}
where $s_k$ and $\dot{s}_k$ are the signal amplitude and rate of change of the amplitude. 
The measurement noise model was characterized by the pixel standard deviation across the sensor in a dark field, and the process noise model was computed using a logarithmic grid search for optimal fit. The denoised signal was determined by the optimal linear estimation of the system~\cite{kalman1960}, and denoised diffuse reflectance was extracted at each time point using an observation matrix on the estimated state:
\begin{equation}
    \hat{R}_d(\vec{x}_i,t_k) = H\hat{\vec{s}}_k
\end{equation}
where $H = [1~0]^T$. Finally, the calibrated attenuation for cross-condition comparison was calculated for each location in each frame:
\begin{equation}
    A(\vec{x}_i,t_k) = -\log(\hat{R}_d(\vec{x}_i,t_k))
    \label{eq:A_xi}
\end{equation}

A spatial correlation filter was performed against the arterial waveform to localize the jugular venous pulse~\cite{amelard2016spatial}. Although arterial waveforms can be extracted directly from the hemodynamic frames, to avoid using the system's own data as input during these validation experiments, a time-synchronized finger photoplethysmography sensor was used to extract the arterial waveform.
Jugular venous pressure is modulated by right atrial pressure, which rises during atrial contraction (ventricular end-diastole) and falls markedly during ventricular systole~\cite{walker1990}.
This contraction-relaxation asynchrony between the atria and ventricles creates pressure waveforms in the jugular venous and arterial circulations that are out of phase~\cite{amelard2017}. 
Specifically, correlation between attenuation signals and the arterial waveform was computed for each tissue location $\vec{x}_i$, producing a venous pulsatility map:
\begin{equation}
    V(\vec{x}_i) = r(A(\vec{x}_i),z) \cdot \mathbbm{1_{\mathbb{R}^-}} \left( r(A(\vec{x}_i),z) \right)
\end{equation}
where $r(A(\vec{x}_i),z)$ is the linear correlation between attenuation signals and the arterial waveform, and $\mathbbm{1_{\mathbb{R}^-}}$ is the indicator function of negative real numbers. Thus, signals that are negatively correlated with the arterial waveform (i.e., out of phase) were identified, along with their associated correlation strength.

Finally, using the venous pulsatility map $V(\vec{x}_i)$, the jugular vein was identified as a contiguous area of strong venous pulsatility along the lateral area of the neck. The calibrated attenuation signals along a $\sim$1~cm length of the vessel track were extracted and spatially averaged for each frame to compute the jugular venous attenuation (JVA) signal. Mathematically:
\begin{equation}
    \text{JVA}(t_k) = \frac{1}{|V|} \sum_{\vec{x}_i \in V} A(\vec{x}_i, t_k)
\end{equation}
where $V$ is the set of pixels within the 1~cm pulsatile venous track. JVA is an attenuation measurement signal derived from the theoretical equations described by Eq.~(1)--(4), and thus can be used to quantify changes in vessel dynamics.

\section{Data Collection and Experimental Design}
\subsection{Participants}
Twenty young healthy adults (10 female, 24$\pm$4~years, 169$\pm$11~cm, 66$\pm$13~kg) with no history of vascular, inflammatory, or thrombotic disease were enrolled. Participants abstained from caffeine, nicotine, alcohol, and heavy exercise for 24~h prior to testing and were fasted for 2~h prior to testing. Participants were asked to consume 2~L of water 2~h prior to testing to ensure a hydrated state. One participant's data were removed due to an insufficient area for imaging, and another participant's LBNP data were removed due to acquisition issues. The study was approved by a University of Waterloo Research Ethics Committee (ORE~\#40394) and all protocols conformed to the Declaration of Helsinki. All participants provided written informed consent prior to testing.

\subsection{Experimental Protocol}
Participants were instrumented with an electrocardiogram (iE33 xMatrix, Philips Healthcare, Andover, USA), continuous arterial blood pressure plethysmogram with estimated stroke volume (SV) via Modelflow (NOVA, Finapres Medical Systems, Enschede, NL), a nasal cannula connected to a capnograph for end-tidal carbon dioxide (\petco; CD-3A CO2 Analyzer, AMETEK Inc., Pittsburgh, PA, USA), and a respiratory belt (Model 0528, Respiratory Effort PVDF Sensor, Braebon, NY, USA). Continuous CVP was measured using a catheter placed in a vein in the right antecubital fossa~\cite{gauer1956continuous}. The vein was punctured using a single-use 20G needle and attached to a pressure monitoring transducer
through a saline-filled polyurethane catheter. 
The pressure transducer was positioned at the level of the right midclavicular line using a laser level to estimate the level of the right heart and manually calibrated at 0, 5, 10, and 15~cmH$_2$O. Participants were tilted to the right to ensure a continuous column of fluid between the right atrium and an antecubital vein. 
Simultaneous vascular ultrasound of the left internal jugular vein was conducted using duplex ultrasound utilizing a 9-3~MHz linear transducer (iE33 xMatrix, Philips Healthcare, Andover, USA). 
LBNP and Valsalva pressures were monitored using a digital manometer. The imaging head was positioned such that it provided orthogonal illumination~\cite{moco2015orthogonal} to the right lateral area of the neck. The system was moved with the participant between conditions to maintain a similar field of view and angle of illumination.

Data were recorded at 1000~Hz (PowerLab, LabChart, version 7.3.7; ADInstruments, Colorado Springs, CO). Hemodynamic frames were captured in uncompressed video format at 60~fps with 16~ms exposure time. The arterial waveform from a finger photoplethysmography sensor was sampled through an analog input into a microcontroller (Arduino Uno) at 100~Hz, transmitted through serial USB to a desktop computer, and down-sampled to 60~Hz to match the video frame rate. When the camera received the signal to start recording, a timer reset command was sent from the camera to the microcontroller through an 8-pin GPIO connector to time-synchronize the start time of the arterial waveform with the camera frames. A hardware trigger was transmitted during image acquisition to PowerLab to synchronize the cardiovascular and imaging acquisitions.

Three maneuvers were performed that systematically changed CVP dynamics: LBNP (reduced venous return from central hypovolemia), HDT (venous congestion through cephalad fluid shift~\cite{martin2016,macias2017book}), and Valsalva maneuver (impaired cardiac filling from increased intrathoracic pressure).
All participants underwent three HDT conditions (0, \textminus3, and \textminus6~\degree) and four LBNP conditions (0, \textminus20, \textminus30, and \textminus40~mmHg), with randomized ordering of the two protocols. Approximately 5~min was spent in each condition. A calibration and 30~s video were acquired during the last minute of each condition. LBNP was terminated if systolic blood pressure fell below 70~mmHg at any point during the collection. HDT and LBNP protocols were separated by at least 5-minutes of supine rest. Participants then voluntarily completed a 15~s Valsalva maneuver to a pressure of 40~mmHg (or the participant's maximum attainable pressure) while breathing into a mouthpiece, with one practice maneuver prior to collection. Data were recorded during the Valsalva, as well as 15~s before (baseline) and after (recovery). To compensate for participant movement during the Valsalva, frames were registered by tracking a salient feature on the neck using a linear dual correlation filter~\cite{henriques2014}.

\subsection{Data Analysis}
Data were analyzed in LabChart (ADInstruments Inc., Colorado, USA). Each physiological signal, including JVA, was averaged across each cardiac cycle using the ECG R~wave times to generate mean beat-by-beat values, which were then averaged across the 30~s acquisition period to produce mean responses for each experimental condition. Mean arterial pressure (MAP) was calculated as beat-by-beat averages of the reconstructed brachial blood pressure waveform (NOVA, Finapres Medical Systems, Enschede, NL).
Baseline and recovery measures during the Valsalva protocol were extracted as 10~s beat-by-beat averages starting at the beginning of the protocol and 5~s after the Valsalva maneuver was stopped, respectively. The peak measure during the Valsalva protocol corresponded with the CVP apex across the respiratory maneuver. The jugular vein was identified by a contiguous negative correlation segment on the right lateral area of the neck in the baseline frames. For each participant, a jugular venous waveform was extracted for each 30~s condition acquisition in the same area in the center of the vessel track, of approximately 1~cm long, identified by a strongly negatively correlated contiguous area close to the clavicle.

Main effects of protocol condition (HDT angle, LBNP pressure level, Valsalva stage) were determined by one-way repeated measures ANOVA followed by planned contrasts between condition to baseline using paired sample t-test when significant main effects were observed. Significant differences were reported at $p<0.05$ with Bonferroni multiple comparison correction. Pearson correlation coefficients were computed to investigate linear relationships between normalized changes in JVA, CVP and SV. Data are presented as mean~$\pm$~standard deviation.

\section{Results}


\begin{figure}
    \centering
    \includegraphics[width=0.45\textwidth]{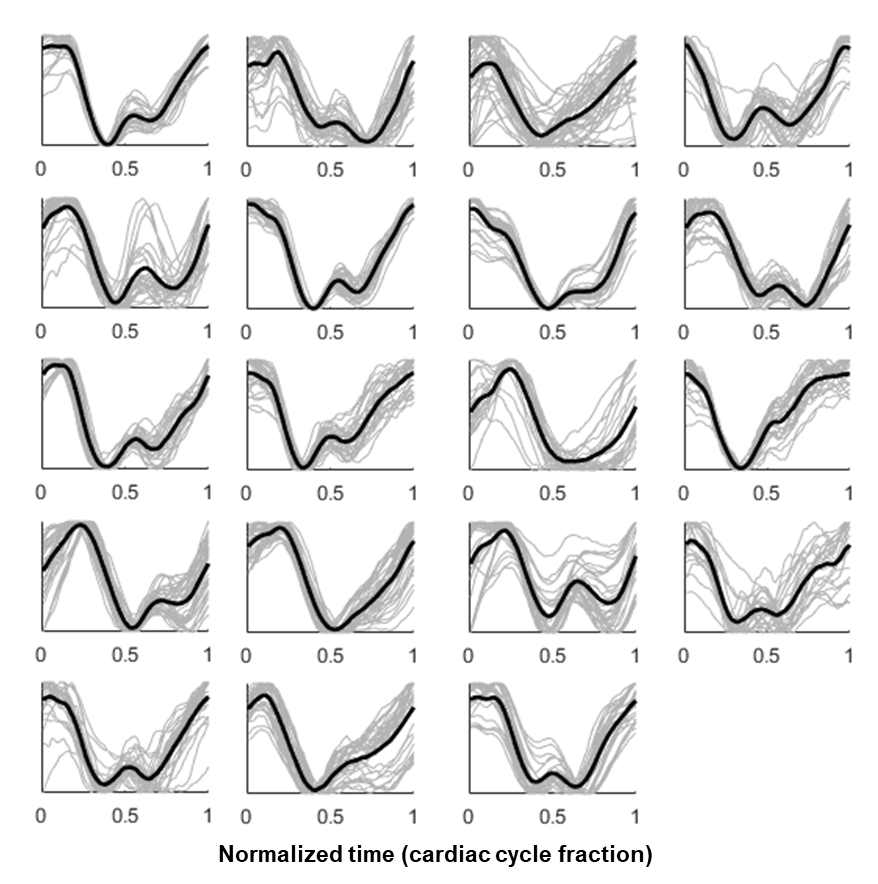}
    \caption{Individual and averaged jugular venous waveforms for each participant while supine, normalized to the cardiac cycle. The start of each signal was time aligned to the ECG R wave.}
    \label{fig:jvps_per_cardiaccyle_hdt0}
\end{figure}




\begin{figure}
    \centering
    \includegraphics[width=0.42\textwidth]{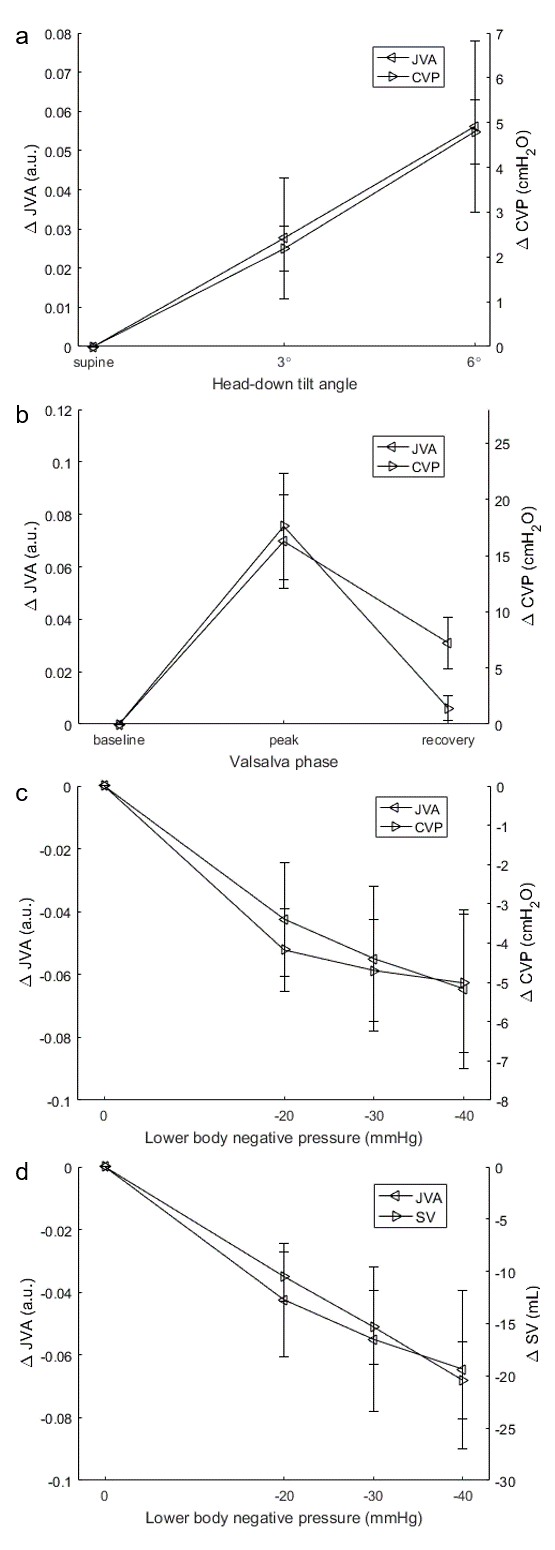}
    \caption{Change in jugular venous attenuation (JVA) compared to central venous pressure (CVP) during (a) head-down tilt, (b) Valsalva, as well as (c) CVP and (d) stroke volume (SV) during lower-body negative pressure. Data are expressed as changes relative to the participant's baseline supine measures. Error bars indicate 95\% confidence intervals. See Table~\ref{tab:results} for statistical comparisons.}
    \label{fig:results_hdt_lbnp}
\end{figure}

\begin{figure}
    \centering
    \includegraphics[width=0.5\textwidth]{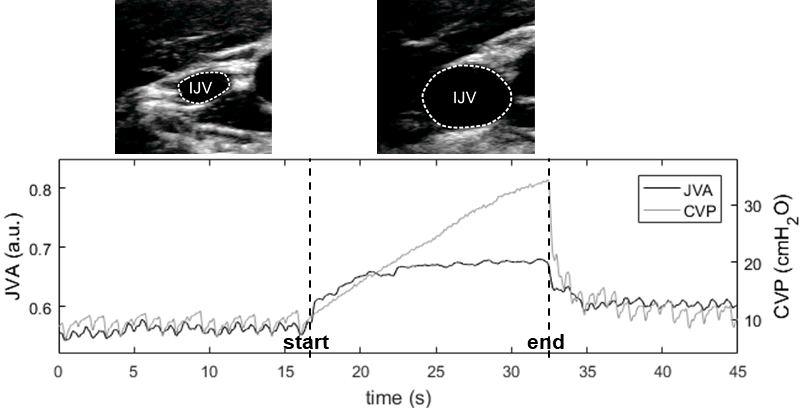}
    \caption{Representative example the dynamics of jugular venous optical attenuation (JVA) and central venous pressure (CVP) during a 15~s Valsalva maneuver in which the participants attempted to exhale through a mouthpiece connected to a pressure transducer to achieve an intrathoracic pressure increase of 40~mmHg. Ultrasound images show the internal jugular vein in supine rest and at peak Valsalva pressure. The plateau in JVA reflects the upper limit of distension of the vein. See Supplementary Video~1 for the full animation.}
    \label{fig:vals_example}
\end{figure}

\begin{table*}[]
\centering
\caption{\textsc{Cardiovascular Response to Three Types of Cardiovascular Stress}}
\label{tab:results}
\begin{tabular}{llccccc}
\hline
\multicolumn{1}{c}{} & \multicolumn{1}{c}{} & \begin{tabular}[c]{@{}c@{}}HR\\ bpm\end{tabular} & \begin{tabular}[c]{@{}c@{}}SV\\ mL\end{tabular} & \begin{tabular}[c]{@{}c@{}}MAP\\ mmHg\end{tabular} & \begin{tabular}[c]{@{}c@{}}CVP\\ cmH$_2$O\end{tabular} & \begin{tabular}[c]{@{}c@{}}JVA\\ a.u.\end{tabular} \\ \hline
\multirow{4}{*}{\begin{tabular}[c]{@{}l@{}}Lower body\\ negative pressure \\(mmHg)\end{tabular}} & 0 & 60 $\pm$ 7 & 79 $\pm$ 15 & 80 $\pm$ 9 & 11.2 $\pm$ 3.5 & 0.56 $\pm$ 0.10 \\
 & \textminus20 & ~64 $\pm$ 9* & ~70 $\pm$ 17* & ~81 $\pm$ 11 & ~~7.0 $\pm$ 3.9* & ~0.51 $\pm$ 0.08* \\
 & \textminus 30 & ~~68 $\pm$ 10* & ~64 $\pm$ 18* & 82 $\pm$ 9 & ~~6.5 $\pm$ 3.9* & ~0.50 $\pm$ 0.07* \\
 & \textminus40 & ~~75 $\pm$ 13* & ~59 $\pm$ 18* & 85 $\pm$ 9 & ~~6.3 $\pm$ 5.0* & ~0.47 $\pm$ 0.05* \\ \hline
\multirow{3}{*}{\begin{tabular}[c]{@{}l@{}}Head-down tilt\\ (tilt angle)\end{tabular}} & 0\degree & 62~$\pm$~8 & 81 $\pm$ 16 & 80 $\pm$ 9 & 11.6 $\pm$ 4.5 & 0.55 $\pm$ 0.09 \\
 & \textminus 3\degree & 61 $\pm$ 9 & 82 $\pm$ 15 & 81 $\pm$ 9 & ~13.8 $\pm$ 4.6* & ~0.58 $\pm$ 0.10* \\
 & \textminus 6\degree & 62 $\pm$ 9 & 80 $\pm$ 15 & ~80 $\pm$ 11 & ~16.4 $\pm$ 4.6* & ~0.61 $\pm$ 0.10* \\ \hline
\multirow{3}{*}{Valsalva} & baseline & ~67 $\pm$ 11 & ~84 $\pm$ 20 & 91 $\pm$ 8 & 11.7 $\pm$ 4.3 & 0.71 $\pm$ 0.23 \\
 & peak & ~~~85 $\pm$ 16* & ~~43 $\pm$ 19* & ~96 $\pm$ 13 & ~~~29.0 $\pm$ 11.6* & ~0.78 $\pm$ 0.22* \\ & recovery & 64 $\pm$ 7 & 83 $\pm$ 19 & 99 $\pm$ 11* & ~13.1 $\pm$ 3.3* & ~0.74 $\pm$ 0.23* \\ \hline
\multicolumn{7}{l}{\begin{minipage}[t]{0.72\textwidth}Values (mean $\pm$ SD) were averaged across the 30~s video acquisition in lower body negative pressure and head-down tilt. Valsalva values show 10~s average 5~s before (baseline) and after (recovery) the maneuver, and during maximal CVP (peak). HR: heart rate; SV: stroke volume; MAP: mean arterial pressure; CVP: central venous pressure; JVA: jugular venous optical attenuation (calibrated). *p$<$0.05 compared to baseline (paired sample t-test) \end{minipage}}
\end{tabular}
\end{table*}

\subsection{Jugular Venous Attenuation Signal Extraction}
Fig.~\ref{fig:jvps_per_cardiaccyle_hdt0} shows per-participant average and individual JVA waveforms during supine baseline. Each signal's start time was set to the ECG R wave, and expressed as a fraction of the cardiac cycle. The biphasic nature and primary subwaves (a, c, x, v, y) of the JVP were visually apparent across participant waveforms. Recognizing the start time as the peak of the R wave (ventricular depolarization), the following pulse wave characteristics were observed: a small rise in pressure during early systole (c wave); a downstroke following the R wave during ventricular contraction from right atrial relaxation and tricuspid valve closure (x wave); an inflection point approximately half way through the cardiac cycle consistent with right atrial filling (v wave) and tricuspid valve opening (y wave); and an upstroke preceding the R wave during atrial contraction (a wave).

\subsection{Acute Central Hypovolemia}
Acute central hypovolemia through graded LBNP resulted in progressively reduced venous return and, accordingly, SV (Fig.~\ref{fig:results_hdt_lbnp}).
Significant main effects were observed in heart rate (HR), SV, CVP, and JVA (Table~\ref{tab:results}).
HR was elevated and SV was depressed during all LBNP levels compared to baseline. These changes in SV are indicative of decreased cardiac filling pressures due to the translocation of blood to the lower extremities. This was corroborated by CVP, which decreased significantly from baseline (11.2$\pm$3.5~cmH$_2$O), with the largest decreases occurring on initial \textminus20~mmHg LBNP (7.0$\pm$3.9~cmH$_2$O) followed by smaller absolute decreases in \textminus30~mmHg (6.5$\pm$3.9~cmH$_2$O) and \textminus40~mmHg (6.3$\pm$5.0~cmH$_2$O) LBNP. A similar pattern of JVA changes was observed, with significant decreases in all grades compared to baseline (0.56$\pm$0.10~a.u.), the largest occurring upon LBNP onset (0.51$\pm$0.08~a.u.) followed by smaller absolute decreases in \textminus30~mmHg (0.50$\pm$0.07~a.u.) and \textminus40~mmHg (0.47$\pm$0.05~a.u.). JVA exhibited strong positive linear correlation (median, interquartile range) to CVP (r=0.85, [0.72, 0.95]) and SV (r=0.85, [0.76, 0.92]) during LBNP.

\subsection{Venous Congestion}
HDT induced a cephalad fluid shift through hydrostatic pressure gradient toward the head, resulting in increased venous pressure (Fig.~\ref{fig:results_hdt_lbnp}).
There were no significant main effects of central cardiovascular variables (HR, SV, MAP) with increased HDT compared to supine baseline (see Table~\ref{tab:results}).
Significant main effects were observed in CVP and JVA.
CVP increased significantly from baseline (11.6$\pm$4.5~cmH$_2$O) during \textminus3$\degree$ (13.8$\pm$4.6~cmH$_2$O) and \textminus6$\degree$ (16.4$\pm$4.6~cmH$_2$O) HDT, with a concomitant significant increase in JVA from supine (0.55$\pm$0.09~a.u.) to \textminus3$\degree$ (0.58$\pm$0.10~a.u.) and \textminus6$\degree$ (0.61$\pm$0.10~a.u.) HDT, signifying increased optical absorption with pressure-induced venous dilation. JVA exhibited strong positive linear correlation (median, interquartile range) to CVP during HDT (r=0.94, [0.84, 0.99]).

\subsection{Impaired Cardiac Filling}
Venous return was experimentally impeded through a Valsalva maneuver. Fig.~\ref{fig:vals_example} shows a representative example of the effect of Valsalva on JVA and CVP signals. Following normal baseline respiration, Valsalva strain resulted in an increase in intrathoracic pressure causing an increase in external pressure on the heart and thoracic blood vessels. Compression of the superior vena cava impedes venous return and results in distended IJV and increases in superior vena cava and central venous pressures~\cite{attubato1994}. This duality was demonstrated by significantly lower SV at peak (43$\pm$19~mL) compared to baseline (84$\pm$20~mL), from reduced venous return, and significantly higher CVP at peak (29.0$\pm$11.6~cmH$_2$O) compared to baseline (11.7$\pm$4.3~cmH$_2$O; see Table~\ref{tab:results} and Fig.~\ref{fig:results_hdt_lbnp}). A similar significant increase in JVA was observed from baseline (0.71$\pm$0.23~a.u.) to peak (0.78$\pm$0.22~a.u.), consistent with jugular venous distension, which reached a plateau at high pressure values when the jugular vein reached its upper limit of distension, as seen in heart failure~\cite{simon2018valsalvahf}. Both CVP and JVA increased with continued fluid congestion until strain release, and exhibited marginally higher levels than baseline during recovery (CVP:~13.1~vs.~11.7~cmH$_2$O; JVA:~0.74~vs.~0.71~a.u.). JVA exhibited strong positive linear correlation (median, interquartile range) to CVP during the baseline-peak-recovery phases (r=0.94, [0.85, 0.99]).

\subsection{Linearity Across Combined Protocols}
\begin{figure}
    \centering
    \includegraphics[width=0.45\textwidth]{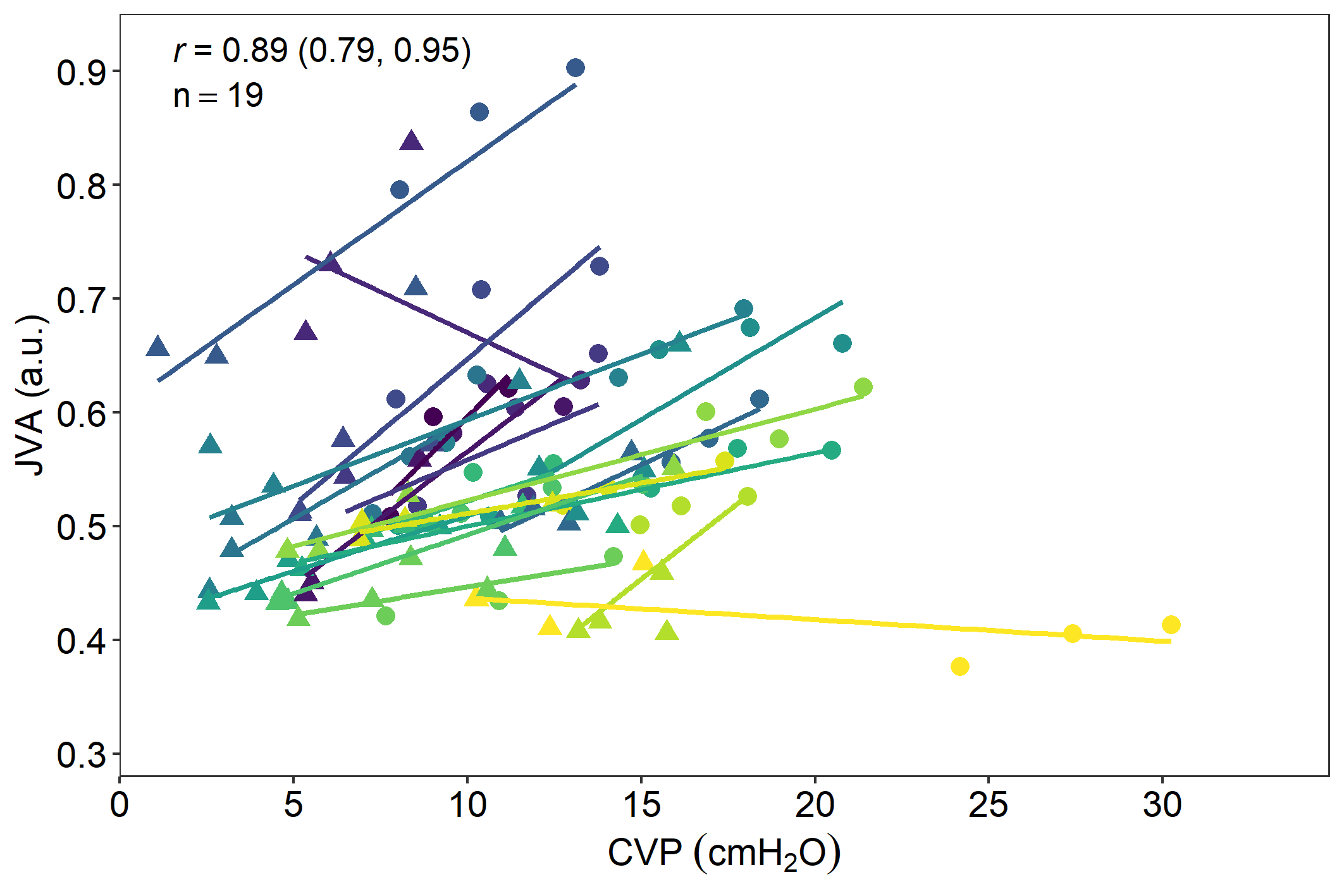}
    \caption{Association between jugular venous attenuation (JVA) and central venous pressure (CVP) for all participants across all head-down tilt (HDT; circles) and lower body negative pressure (LBNP; triangles) protocols. Each participant is represented by a different color. The correlation coefficient distribution across the sample is reported as median (interquartile range).}
    \label{fig:results_allhdtlbnp}
\end{figure}

The linearity of the system across a wide range of CVP values was assessed by linear regression of JVA against CVP across all HDT and LBNP protocols (Fig.~\ref{fig:results_allhdtlbnp}). JVA exhibited strong within-participant positive linear correlation (median, interquartile range) to CVP across all protocols (r=0.89, [0.79, 0.95]). Inter-participant intercept and slope differences are attributed to differences in relative contributions of venous blood and tissue absorption/scattering due to inter-participant differences in tissue composition.

\section{Discussion}
In this work, we proposed a non-contact, sub-surface optical imaging system to monitor changes in jugular venous dynamics. We have shown that JVA closely tracks the effects of increases and decreases in CVP during hemodynamic stresses induced by LBNP, HDT and Valsalva.  Three physiological protocols were designed to test multiple mechanisms of CVP modulation: (1) reducing venous return through LBNP; and (2) increasing venous pressure through venous congestion in HDT; and (3) restricting venous return by increasing intrathoracic pressure with a Valsalva maneuver. Taken together, the span of effects of these three protocols have clinical and physiological significance, as discussed below. These observations support the hypothesis that pressure-induced changes in jugular venous size can be monitored in a non-contact way through calibrated optical hemodynamic imaging. The current clinical standard for monitoring CVP involves invasive right heart catheterization~\cite{wong2017canadian}, which limits the scope of monitoring primarily to advanced stage cases~\cite{tibby2003monitoring}. Non-invasive optical imaging of the jugular venous response to pressure changes would enable a wider range of monitoring applications, both clinically and physiologically.

Non-contact optical imaging methods for extracting hemodynamic signals, or summary measures such as HR, are relatively new technologies~\cite{sun2015review}. Instrumentation varies between studies, but many comprise some combination of a camera with or without a controlled light source. Relying on the same optical theory as finger photoplethysmography, these systems produce unitless measurements, and thus are not suitable for longitudinal comparison in their basic form. Diffuse optical imaging and spectroscopy methods are able to quantify absorption per unit length pathlength through calibration procedures~\cite{cuccia2009,bevilacqua2000dosi}, but are typically restricted in geometry and assumptions about tissue homogeneity. We combined the two approaches to yield a calibrated non-contact optical imaging method for comparing changes in optical attenuation from venous dilation by effectively determining the per-pixel incident illumination by normalizing inhomogeneities in the optical system and resulting from the surface profile. In these experiments, we used an arterial waveform from a time-synchronized finger sensor in order to minimize potential bias of using non-validated imaging data for identifying jugular venous pulsatility. In future work, the arterial waveform may be extracted directly from the hemodynamic frames, eliminating the need for time-synchronizing circuitry and a contact-based sensor.

LBNP causes a translocation of central blood volume to the lower body vascular compartments without altering gravitational or muscular loading effects, and is an experimental model for cardiovascular response to orthostatic stress and acute hemorrhage~\cite{cooke2004lbnphemorrhage,alian2014}. Our experimental design simulated the equivalent of approximately 500--1000~mL of blood loss across the chosen pressure levels~\cite{hanson1998}. This central hypovolemia reduced venous return leading to activation of cardiopulmonary baroreceptors and reduction in stroke volume that reduces arterial pulse pressure causing elevated HR, sympathoexcitation and release of arginine vasopressin to reflexively maintain arterial and cerebral perfusion pressures~\cite{norsk1993lbnp}. Our data tracked the effects of greater LBNP with reduced CVP and commensurate JVA. In some participants, CVP was close to zero during the last stage of LBNP. MAP remained unchanged after 4~min of exposure to each level, indicating an effective cardiovascular response of vasoconstriction and increased HR. Monitoring arterial pressure may lead to missed early biomarkers of shock~\cite{wilson2003trauma}, whereas venous pressures may be more indicative of early signs of hypovolemia~\cite{alian2014}. Thus, optical monitoring of jugular venous changes may enable enhanced monitoring in trauma.

Venous congestion and stasis during spaceflight results from the removal of the gravitational force which modulates venous return. In comparison to on Earth, a cephalad fluid shift moves venous blood away from the lower limbs toward the heart and head, and has been associated with increased intracranial pressure and jugular vein thrombosis in space~\cite{lawley2017icp,marshallgoebel2019}. HDT has been extensively used as an Earth-based analog to study cardiovascular responses to spaceflight, with 6$\degree$ being widely used as a model of 0~G~\cite{hargens2016hdt}. The resulting jugular distension can be visually seen on astronauts aboard the International Space Station, and is consistent with increases in JVA observed in this study. Exposure to 24~h simulated spaceflight microgravity has shown that intracranial pressure levels are chronically elevated above those observed during 90$\degree$ seated position on Earth~\cite{lawley2017icp}, which has implications in ocular remodeling in space~\cite{zhang2018}. A mismatch between intracranial and intraocular pressure is one of the primary hypotheses behind visual acuity impairments in space~\cite{mader2011sans}. Non-intrusive continuous measurements of jugular venous volume and waveform over an entire mission may enable identification of abnormal responses to spaceflight, or evaluation of the effectiveness of countermeasures designed to limit the severity of spaceflight associated neuro-ocular syndrome.

Restricted venous return is present is right heart failure, where inefficient cardiac function leads to fluid overload. In congestive heart failure, reduced cardiac function results in central fluid accumulation with elevated CVP. Jugular venous distension is associated with prognosis of hospitalization and disease progression in heart failure~\cite{drazner2001}. Thus, a large emphasis is placed on managing, and therefore monitoring, fluid levels~\cite{gelman2018physiologic}. Since jugular venous distension correlates well to right atrial pressure and left ventricular filling pressures~\cite{davison1974,simon2018valsalvahf}, we used a Valsalva maneuver to systematically increase intrathoracic and cardiac pressures. Changes in jugular venous cross-sectional area through ultrasound assessment are a positive predictor of heart failure status due largely to elevated right atrial pressure~\cite{simon2018valsalvahf,pellicori2015valsalvahf}. In our healthy participant sample, we observed significant changes in JVA during the Valsalva strain compared to baseline, indicating a measured difference in jugular distension, and suggesting that it may be a clinically relevant non-contact alternative for heart failure monitoring.

Regression analysis of JVA against all HDT and LBNP protocols was performed without constraining slope or intercept to model between-participant differences in vessel compliance as well as tissue composition, and thus static optical properties. Factors such as vessel depth, melanin distribution, and surrounding tissue composition impact light attenuation~\cite{jacques2013,martelli2016photondepth}. By assuming these other factors to be constant, this setup models the relative contributions of venous blood and tissue absorption/scattering. As a result, different regression slopes and intercepts were found between participants, with an overall strong positive linear relationship between CVP and JVA. A negative slope was found in 2/19 participants, indicating decreased attenuation with increased CVP, likely resulting from changes in tissue composition across the protocols. In one of these participants, LBNP was terminated early when arterial blood pressure reached presyncope levels, which is often accompanied by marked pallor and changes in subcutaneous blood flow~\cite{benditt1995syncope}. Further developments are needed for situations in which the assumptions stated in this study may not hold.




In practice, in order to enable non-contact monitoring, this system requires imaging calibration in the monitoring configuration, and would thus be most effectively used in a point-of-care environment. Practical considerations, such as camera positioning, calibration, and lighting, may present challenges for continuous monitoring. Wearable sensors that can be affixed to the patient's skin, such as pressure and strain sensors~\cite{trung2016sensors}, may provide alternative solutions for continuous monitoring, but capturing spatial information for visually identifying and assessing the jugular vein may be challenging, and consistent sensor placement poses a challenge for follow-up assessment. Clinical imaging tools have been helpful for identifying vessel structure that cannot be visually discerned for guiding treatment~\cite{miyake2006vein}. Ultrasound can be used to assess jugular vein cross-sectional area in heart failure~\cite{pellicori2015valsalvahf,simon2018valsalvahf}, however this requires a trained technician to maintain probe contact with care as to not collapse or distort the compliant jugular vein. Although the proposed optical imaging requires calibration and limited motion, it may provide added benefits in settings where wearable sensors are unsuitable.

The primary limitation of this work is the assumption that all other physiological and environmental systems remain constant across conditions, and that the change in optical attenuation was solely due to changes in blood volume. This was achieved by performing the three protocols in a single data collection session, where environmental and physiological conditions could be controlled. However, during longitudinal monitoring, additional factors that could alter tissue optical properties must be explicitly modeled, such as blood oxygen saturation, hematocrit, and skin blood volume. We also acknowledge that CVP was used to validate observed changes in JVA, which is a measurement of blood volume in the vein, and could be influenced by venous compliance. Exceptionally high pressures may result in a reduced volume response. Thus, the generalizability of these results to changes in already high venous pressures as well as changes in venous tone requires further research with affected populations.

\section{Conclusion}
In this paper, we proposed a non-contact optical imaging system for assessing changes in jugular venous optical attenuation resulting from altered central venous pressure. Comparison of optical attenuation of the jugular vein in different imaging configurations was performed using a proposed surface profile calibration and signal denoising pipeline. Using a time-synchronized arterial waveform, the jugular venous waveform was extracted and compared in three physiological protocols: central hypovolemia, venous congestion, and impaired cardiac filling. Results showed that changes in jugular venous optical attenuation strongly correlated with changes in central venous pressure, as well as reduced stroke volume during central hypovolemia. These results suggest that non-contact optical imaging may be used for assessing clinically relevant hemodynamic conditions without the need for invasive catheterization.

\bibliographystyle{IEEEtran}
\bibliography{bibliography}

\newpage

\end{document}